\documentclass[english,aps,pra, twocolumn, floatfix, showpacs, 10pt]{revtex4-1}
\usepackage{amsfonts}
\usepackage{amsmath}
\usepackage{graphicx}
\usepackage{color}
 \usepackage{hyperref}
 \usepackage{natbib}
\begin{document}
\title {Excitation spectra of a Bose--Einstein condensate\\with an angular spin--orbit coupling}
\author{Ivana Vasi\'c}
\affiliation{Scientific Computing Laboratory, Center for the Study of Complex Systems, Institute of Physics Belgrade, University of Belgrade, Pregrevica 118, 11080 Belgrade, Serbia}
\author{Antun Bala\v z}
\affiliation{Scientific Computing Laboratory, Center for the Study of Complex Systems, Institute of Physics Belgrade, University of Belgrade, Pregrevica 118, 11080 Belgrade, Serbia}

 \begin{abstract}
A theoretical model of a Bose--Einstein condensate with angular spin--orbit coupling has recently been proposed and it has been established that a half--skyrmion  represents the ground state in a certain regime of spin--orbit coupling and interaction.  Here we investigate low--lying excitations of this phase by using the Bogoliubov method and numerical simulations of the time--dependent Gross--Pitaevskii equation.
We find that a sudden shift of the trap bottom results in a complex two--dimensional motion of the system's center of mass that  is markedly different from the response of a competing phase, and comprises two dominant frequencies. Moreover, the breathing mode frequency of the half--skyrmion is set by both the spin--orbit coupling and the interaction strength, while in the competing state it takes a universal value. Effects of interactions are especially pronounced at the transition between the two phases.
 \end{abstract}
\maketitle
 
 \section{Introduction}
 
Experimental realization of an effective spin--orbit coupling in ultracold atom systems \cite{Spielman1, Spielman2, Zhangexp, Khamehchi, Cheuk, Li} has allowed for new quantum phases to be explored. Bosonic systems with spin--orbit coupling are interesting as they have no direct analogues in condensed matter systems and provide a new research playground. Different types of coupling based on atom--light interactions have been considered, e.~g.~Raman induced (as realized in the current experiments) and Rashba--type \cite{Zhaireview}. Only recently bosonic systems with two--dimensional spin--orbit coupling have become experimentally available \cite{Wu}.
Ground--state phase diagrams that comprise a plane--wave, stripe and non--magnetic condensed phase have been predicted and probed \cite{Stanescu, Zhai, Ho, YunLi0, Ji}. 
Another type of condensate, a half--quantum vortex, is expected for harmonically trapped bosons with Rashba coupling \cite{Sinha, Hu2, Ramachandhran}. A substantial progress in the field has been summarized in Refs.~\cite{Galitski, Zhaireview}.
As a further extension of these ideas,
in the very recent papers \cite{Hu, DeMarco, Qu, Sun, ChenL, ZhuDongPu}, a theoretical model of bosons with the coupling of spin and angular--momentum has been introduced. From the experimental side, the proposal involves two copropagating Laguerre--Gauss laser beams that carry angular momentum and couple two internal states of bosonic atoms. 
% In the weakly interacting limit, below a threshold coupling strength, the ground--state is found to be a half-skyrmion configuration. By increasing either interaction or spin--orbit coupling strength, a first order phase transition occurs and the ground state is then given by a pair of vortices with opposite circulation. 

Since the first experimental realization of Bose--Einstein condensation, collective modes have been used to probe the macroscopic quantum state and to relate measurements to theoretical predictions \cite{Giorgini}. Collective modes can reveal important information about system properties, such as role of interactions or quantum fluctuations.
Experimentally, breathing mode and dipole mode excitations introduced through a quench of the harmonic trap are routinely accessible with great precision thus providing an indispensable tool for probing the properties of a Bose--Einstein condensate.
Along these lines, collective modes of  bosons with the Raman--induced spin--orbit coupling have already been measured
\cite{Zhangexp, Qu, Khamehchi, Hamner}. In the literature, several theoretical calculations of collective modes for different types of spin--orbit coupling are available \cite{Bijl, Zhangmf, YunLi, Zheng, Chen, Martone, Ramachandhran, YunLi2, Ozawa, Price, Mardonov, Mardonov2}. In contrast to usual, harmonically trapped systems, spin--orbit coupled systems exhibit the absence of the Galilean invariance and as a consequence, the Kohn theorem no longer applies \cite{Zhaireview}. Another hallmark of these systems is that the motion in real space is coupled with spin dynamics. 

In this paper we investigate collective modes of bosons with angular spin--orbit coupling, that have not been addressed so far, and show that the two competing ground states can be directly distinguished according to their response to standard quenches of  the underlying harmonic trap. The paper is organized as follows: In Sec.~II we introduce the basic model and discuss its excitations in the non--interacting limit. In Sec.~III we briefly describe methods that we use and summarize the ground--state phase diagram in the limit of weak interactions \cite{Hu}. Finally, in Sec.~IV we address breathing--mode and dipole mode excitations of the two relevant phases and in Sec.~V we present our concluding remarks.
 
 \section{Non--interacting model}
 
In recent Refs.~\cite{Hu, DeMarco, Qu, Sun} the following Hamiltonian for a two--component bosonic system has been introduced:
\begin{equation}
 H_0 = \left(\frac{p^2}{2}+\frac{ r^2}{2}  \right) \mathcal{I}_2+\frac{\Omega^2 r^2}{2} \left(\begin{array}{cc}
                                                                                                                        1 & e^{-2 i \phi}\\
                                                                                                                        e^{2 i \phi} & 1
                                                                                                                       \end{array}\right),
                                                                                                                       \label{eq:h0}
\end{equation}
where $\mathcal{I}_2$ is a $2\times 2$ identity matrix and the effective spin $1/2$ comes from the two bosonic components involved. The system is assumed to be effectively two--dimensional (tightly trapped in the longitudinal direction) and the value of $\Omega$ is proportional to the intensity of the applied Laguerre--Gauss laser beam. The last, $\phi$--dependent term, where $ \phi$ is the polar angle, provides the coupling between the spin and angular momentum, as can be explicated by using a proper unitary transformation \cite{DeMarco}.  
\begin{figure}[!t]
\includegraphics[width=8cm]{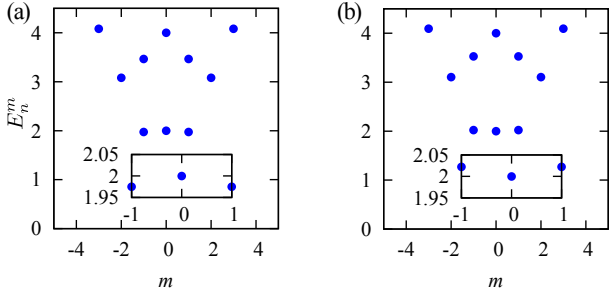}
\caption{Spectrum $E_n^m$ of Hamiltonian (\ref{eq:h0}) for: (a) $\Omega = 3.2$ and (b) $\Omega = 3.5$.}
\label{Fig:Fig0}
\end{figure}
We have assumed that the two lasers carry a  unit of angular momentum in the opposite rotational directions. In Eq.~(\ref{eq:h0}) and in the following we use harmonic oscillator scales of the kinetic energy and the trap ${ p}^2/2m+m\omega^2 r^2/2$ as our units: the energy is expressed in terms of $\hbar \omega$, the unit length is the harmonic oscilator length scale $\sqrt{\hbar/m\omega}$, where $m$ is the atomic mass, the unit momentum is $ \sqrt{\hbar m\omega}$, and the time scale is given by $\omega^{-1}$. The frequency $\Omega$ and all excitation frequencies are expressed in units of the harmonic oscillator frequency $\omega$.

From the commutation relation $\left[J_z,   \mathcal{H}_0\right] =0$, where $J_z = L_z \otimes \mathcal{I} + \mathcal{I} \otimes \sigma_z$ is the $z$ component of the total angular momentum, it follows that the non--interacting eigenstates can be written in the form
 \begin{equation}
 \phi_m(r, \phi)=\frac{e^{i m \phi}}{\sqrt{2\pi}}\left(\begin{array}{c}
                                                                                              f_m(r) e^{-i \phi}\\
                                                                                              g_m(r) e^{i \phi}
                                                                                             \end{array}\right),
                                                                                             \label{eq:phim}
\end{equation}
where $m$ as an eigenvalue of $J_z$ takes integer values and $r$ is the radial coordinate. By numerical calculation \cite{Hu} it has been shown that the ground state moves from the $m=1$ into the $m=0$ subspace at $\Omega_c \approx 3.35$. The $m=1$ ground state exhibits a non--trivial spin texture that can be characterized by a topological number (a winding number of the spin vector). This state is called half--skyrmion and is degenerate, i.~e.~it has the same energy as the ground state in the $m=-1$ subspace. The $m=0$ states comprises two vortices of opposite circulation.

We investigate excitations above the half--skyrmion and $m=0$ ground state, first at a single--particle level. 
The spectrum of the Hamiltonian (\ref{eq:h0}) is shown in Fig.~\ref{Fig:Fig0}(a) for $\Omega=3.2$ and in Fig.~\ref{Fig:Fig0}(b) for $\Omega=3.5$.  In the first case, for $\Omega = 3.2 < \Omega_c$ the ground state $m=1$ is doubly degenerate and the lowest $m = 0$ state is close in energy, $E_0^{m=0}-E_0^{m=1}\approx 2.5 \times 10^{-2}$. For $\Omega = 3.5 > \Omega_c$ the ground state corresponds to $m=0$. In the following we will probe some features of these spectra by applying two experimentally relevant types of perturbations to a selected ground state.

To induce a breathing mode, we  perturb the trap strength
\begin{equation}
 H_{\mathrm{pert}}  = H_0 +  \eta  \frac{r^2}{2}  \mathcal{I}_2.
 \label{eq:pertbm}
\end{equation}
\begin{figure}[!t]  
\includegraphics[width=8cm]{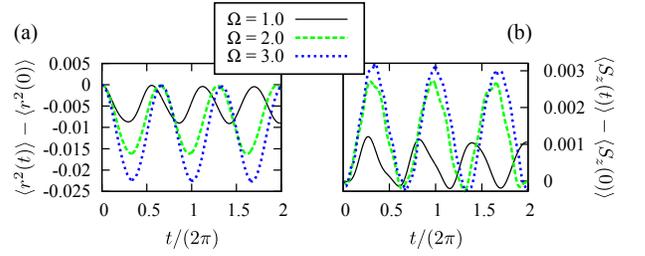}
\caption{Breathing mode oscillations of half--skyrmion state, evidenced by  (a) $ \langle r^2(t) \rangle - \langle r^2(0)\rangle$ and (b) $ \langle S_z(t) \rangle -\langle S_z(0) \rangle$. Motion is induced by changing harmonic trap potential as $\frac{r^2}{2} \rightarrow 1.01\frac{r^2}{2} $.}
\label{Fig:Figbm0}
\end{figure}
From the time--dependent Schr\"odinger equation,
\begin{equation}
 i \frac{\partial}{\partial t} \left(\begin{array}{c} \psi_1(t)\\ \psi_2(t) \end{array} \right) =  H_{\mathrm{pert}}  \left(\begin{array}{c} \psi_1(t)\\ \psi_2(t) \end{array} \right),
\end{equation}
we calculate the time evolution of the width of the probability distribution,
\begin{equation}
 \langle r^2(t) \rangle = \int_0^{2\pi} d\phi \int_0^{\infty} dr \,r^3 \left(|\psi_1(t)|^2+|\psi_2(t)|^2\right),
\end{equation}
as well as the spin dynamics captured by
\begin{equation}
 \langle S_z(t) \rangle = \frac{1}{2}\int_0^{2\pi} d\phi \int_0^{\infty} dr\, r \left(|\psi_1(t)|^2-|\psi_2(t)|^2\right).
\end{equation}
When changing the trap strength $\eta$ in the Hamiltonian (\ref{eq:pertbm}), we couple only states with the same value of $m$. In the limit of vanishing $\Omega$, the breathing mode frequency is $\omega_B = 2$. By increasing $\Omega$, while staying in a half--skyrmion state, we find that the breathing mode frequency decreases down to $\omega_B \approx 1.5$ at the transition point, Fig.~\ref{Fig:Figbm0}(a). Oscillations in the system size are accompanied by an oscillatory spin dynamics, as shown in Fig.~\ref{Fig:Figbm0}(b).
%The main result is given in Fig.~\ref{Fig:Fig0}b where we also anticipate results of later sections. 

In the $m = 0$ subspace, by subtracting and summing the two coupled eigenequations,  we find that the eigenproblem reduces to two independent harmonic oscillators,
\begin{eqnarray}
 \left(\mathcal{L} + \frac{\left(1+2\Omega^2\right)r^2}{2} \right)\left(f_0(r)+g_0(r)\right) &=& E^{m=0} \left(f_0(r)+g_0(r)\right),\nonumber\\
  \left(\mathcal{L}+ \frac{r^2}{2}\right) \left(f_0(r)-g_0(r)\right) &=& E^{m=0} \left(f_0(r)-g_0(r)\right),\nonumber
\end{eqnarray}
with frequencies $1$ and $\sqrt{1+2\Omega^2}$, and the azimuthal quantum number $1$ in both cases as $\mathcal{L} = -\frac{1}{2r} \frac{\partial}{\partial r}\left(r\frac{\partial}{\partial r}\right)+\frac{1}{2 r^2}$. Hence, the $m=0$ energy levels are linear combinations of $E_n^{m = 0}=2 n$ and $E_n^{m = 0}=2 \sqrt{1+2\Omega^2}\, n, n=1,2,\ldots$. In the region of interest, where $\Omega$ is strong enough, the ground--state energy is exactly $E_0^{m = 0}=2$ with a wave function
\begin{equation}
\phi_0=\frac{1}{\sqrt{2\pi}}\left(\begin{array}{c}
                                                                                              f_0(r) e^{-i \phi}\\
                                                                                              -f_0(r) e^{i \phi}
                                                                                             \end{array}\right),
\end{equation}
which is independent of $\Omega$.
From this analysis it follows that the breathing mode frequency is $\omega_B = 2$, which is a well--known result for harmonically trapped bosons in two dimensions at the classical level \cite{Pitaevskii}. Moreover, it is easy to show that the time evolution according to the perturbed Hamiltonian (\ref{eq:pertbm}) is given by $\phi_0({\bf r}, t) = \left(\begin{array}{c}
                                                                                              f_0(r, t) e^{-i \phi}\\
                                                                                              -f_0(r, t) e^{i \phi}
                                                                                             \end{array}\right) $, leading to $\langle S_z(t)\rangle =0$. Therefore, in this case oscillations in the system size are not followed by oscillations in $\langle S_z(t) \rangle $.

\begin{figure}[!t]
\includegraphics[width=8cm]{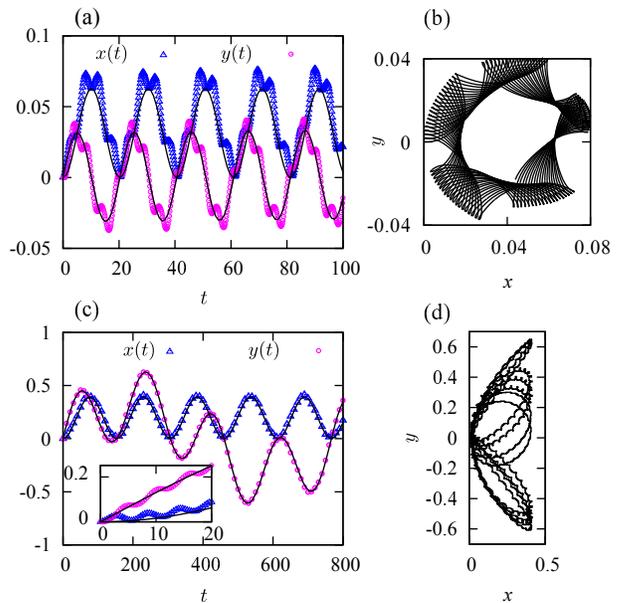}
\caption{Dipole mode oscillations of half--skyrmion state for: $\delta x = 0.02$ at (a), (b)  $\Omega = 2$ and (c), (d) $\Omega = 3.2$. }
\label{Fig:FigdmM1g0}
\end{figure}

To excite a dipole mode, we consider a shift of the trap bottom in $x$ direction,
\begin{equation}
  H_{\mathrm{pert}} = H_0 - \delta x  \frac{r}{2}\left( e^{i \phi} +e^{-i \phi}\right) \mathcal{I}_2,
  \label{eq:pertdm}
\end{equation}
and monitor the motion of the center of mass of the system in that direction,
\begin{equation}
 \langle x(t) \rangle = \int_0^{2\pi} \hspace{-0.3cm} d\phi \frac{\left( e^{i \phi} +e^{-i \phi}\right)}{2}\int_0^{\infty}\hspace{-0.3cm} dr\, r^2 \left(|\psi_1(t)|^2+|\psi_2(t)|^2\right),
\end{equation}
as well as $\langle y(t) \rangle$. In Fig.~\ref{Fig:FigdmM1g0}(a) for $\Omega = 2$ we see that oscillations in $x$ and $y$ directions are coupled and that there are several frequencies involved. 
In Fig.~\ref{Fig:FigdmM1g0}(c) we observe that for $\Omega = 3.2$ even a weak shift of $\delta x = 0.02$, leads to very strong, slow oscillations in $x$ and $y$ directions. 
%This is a consequence of the presence of a low--lying excited state with energy $E_0^{m=0} $. 
On top of this, we also find fast oscillations, as shown in the inset of the same figure. In Figs.~\ref{Fig:FigdmM1g0}(b) and \ref{Fig:FigdmM1g0}(d) we show the resulting complex motion of the center of mass of the system, given by $y(t)$ vs.~$x(t)$. These are all very distinct features not present in the conventional harmonically trapped system, where the same perturbation excites the Kohn mode -- an oscilation with the trap frequency along $x$ axis. In the following we discuss the origin of the complex dynamics.

First we note that the perturbation introduced in the Hamiltonian (\ref{eq:pertdm}) couples the initial $m=1$ ground state with excited states corresponding to other eigenvalues of $J_z$, e.g.~$\int_0^{\infty} dr\, r \int_0^{2 \pi} d\phi\, \phi^*_1(r)  H_{\mathrm{pert}} \phi_0(r)\neq 0$. In general, this effect may lead to the time dependent expectation value $\langle \psi(t)|J_z|\psi(t)\rangle=\langle J_z(t) \rangle $. From the Heisenberg's equations of motion $ i \frac{d J_z(t)}{d t} = \left[J_z, H_{\mathrm{pert}}\right]$ and from the commutation relation $\left[J_z, x \otimes \mathcal{I}_2\right] = i y \otimes \mathcal{I}_2$, we directly obtain that oscillating $\langle J_z(t) \rangle $ implies a motion in $y$ direction
\begin{equation}
 \langle y(t) \otimes \mathcal{I}_2\rangle =  -\frac{1}{\delta x} \frac{d \langle J_z(t) \rangle}{d t}.
 \label{eq:ymotion}
\end{equation}

Now we discuss the emerging oscillation frequencies. In first order of perturbation theory, we would expect the dominant coupling of $m=1$ with $m=0$ and $m=2$ eigenstates, providing the two frequencies:
\begin{equation}
 \omega_D^L = E_0^{m=0}-E_0^{m=1},\quad \omega_D^H = E_0^{m=2}-E_0^{m=1}.
\end{equation}
However,  due to the degeneracy of the states $m=-1$ and $m=1$, the $m=-1$ state has to be taken into account as well. The lowest frequencies can be described by using the perturbation theory for degenerate states presented in Appendix \ref{Sec:Appendix1}. Within this approach we find that the excitation frequencies are
%The eigenvalues of the previous matrix are $E_0^{m=1}, (E_0^{m=0}+E_0^{m=1})/2\pm \sqrt{{\omega_D^L}^2+2 \left(\delta x I_{10}\right)^2}/2 $ 
%leading to the excitation frequencies
\begin{eqnarray}
 \omega_1 &=& \sqrt{{\omega_D^L}^2+2 (\delta x I_{10})^2},\\
 \omega_{2,3} &=&\left|\frac{\omega_D^L}{2}\pm \frac{1}{2}\sqrt{{\omega_D^L}^2+2 (\delta x I_{10})^2}\right|,
\end{eqnarray}
%All calculational details are given in Appendix \ref{Sec:Appendix1}.
 together with $\omega_D^H $.  Obviously, the excited frequencies are amplitude--dependent, and when $\omega_D^L$ is low, i.e.~close to the transition point, the contribution of the term proportional to the trap displacement $\delta x$ is significant. This is another difference with respect to a standard harmonically trapped system. It arises due to the fact that by shifting the trap bottom, while keeping the term proportional to $\Omega^2$ unchanged in the model (\ref{eq:h0}), we lower the symmetry of the model and modify its energy levels. In the regime $\omega_D^L\rightarrow 0 $ it turns out that $\omega_1$ corresponds to oscillations in $x$ direction, while both $\omega_2$ and $\omega_3$ represent the motion in $y$ direction. Results of the analytical calculation, Eqs.~(\ref{eq:xoft}) and (\ref{eq:yoft}) from Appendix \ref{Sec:Appendix1}, are given by the black solid lines in Figs.~\ref{Fig:FigdmM1g0}(a) and \ref{Fig:FigdmM1g0}(c) and capture the low--lying frequencies or long--time dynamics quite well.

\begin{figure}[!t]
\includegraphics[width=8cm]{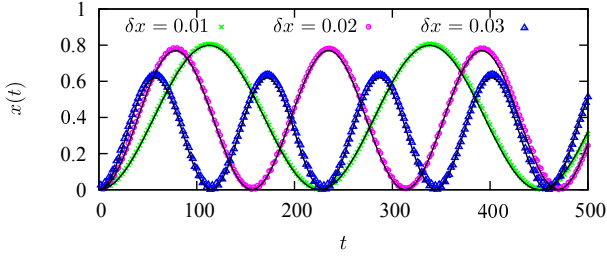}
\caption{Dipole mode oscillations for $\Omega = 3.5$, starting from $m = 0$ ground state with different trap displacements $\delta x$. Black solid lines are results of the analytical calculation.}
\label{Fig:FigdmM0g0}
\end{figure}
The response of a vortex--antivortex pair to the sudden shift of the trap is shown in Fig.~\ref{Fig:FigdmM0g0}. In this case, the perturbation couples the initial $m=0$ state symmetrically to excited states $\pm m$. Thus, $\langle J_z(t)\rangle=0 $ and the center of mass only oscillates in $x$ direction. The two involved frequencies are
\begin{equation}
 \omega_1 = \sqrt{{\omega_D^L}^2+2 (\delta x I_{10})^2},\quad \omega_D^H = E_1^{m=1}-E_0^{m=0}.
\end{equation}
For $\Omega = 3.5$, we have $\omega_D^L \approx 2.2 \times 10^{-2}$ and the increase of the excited frequency with the shift $\delta x$ is clearly observable in the long--time dynamics, see Fig.~\ref{Fig:FigdmM0g0}.

\begin{figure}[!t]
\includegraphics[width=\linewidth]{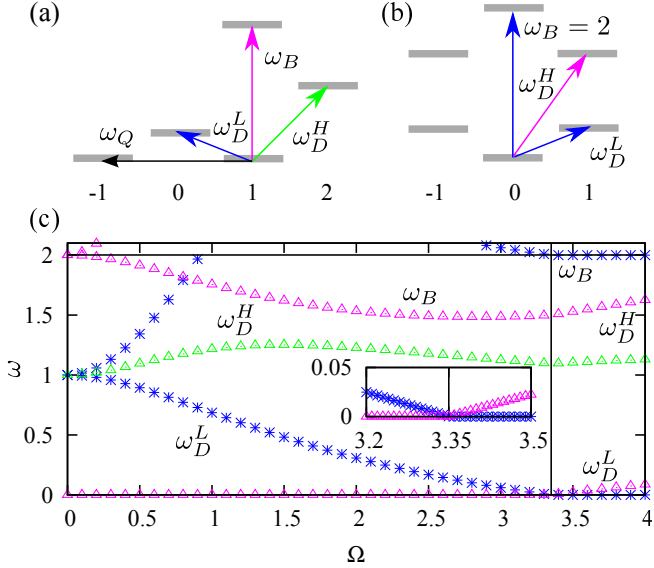}
\caption{Breathing--mode and dipole mode excitations at $g = 0$ of (a) $m = 1$ and (b) $m = 0$ ground state. (c) Energy of excited states as a function of $\Omega$.}
\label{Fig:Figg0}
\end{figure}

Results of this section are summarized in Fig.~\ref{Fig:Figg0}, where we see that at the transition point, $\Omega \approx 3.35 $, $\omega_D^L$ becomes gapless;
$\omega_B$ of the $m=1$ state decreases from $\omega_B =2$ down to $\omega_B \approx 1.5 $ and turns into $\omega_D^H$ of $m=0$ state. On the other hand, $\omega_B = 2$ on top of the $m=0$ ground state  is unaffected by $\Omega$.
We also keep in mind that, due to the degeneracy of the half--skyrmion, below the transition point we have a gapless quadrupole mode $\omega_Q = E^{m=-1}_0-E^{m=1}_0 = 0$ that indirectly affects dipole mode oscillations. For completeness, we note that the frequency $\omega_D^H $ of the half--skyrmion turns into a quadrupole mode of $m=0$ state, but this excitation does not play an important role in the remaining discussion.

\section{Weak interactions}
Now we consider weak spin--symmetric interactions, which are approximated by a contact potential \cite{Zhaireview, Hu, DeMarco}. The total Hamiltonian takes the form
\begin{equation}
 \mathcal{H} = \int d{\bf r}\,\left[ \Psi^{\dagger}({\bf r})  H_0 \Psi({\bf r})+\frac{g}{2} \sum_{a,b=1}^2\,\Psi_a^{\dagger}({\bf r}) \Psi_b^{\dagger}({\bf r}) \Psi_b({\bf r})\Psi_a({\bf r})\right],
 \label{eq:inth}
\end{equation}
where $\Psi({\bf r}) $ is a two--component spinor.
Without interactions, the ground state of many bosons is degenerate for $\Omega < \Omega_c$ as there are different possibilities to accommodate atoms into the two lowest degenerate noninteracting states. In general, the degeneracy of noninteracting eigenstates makes the occurrence of Bose--Einstein condensation more subtle \cite{Stanescu, Moller}. In the case that we consider, it turns out that weak interactions promote condensation \cite{Zhaireview}, as it is energetically favorable for the particles to condense into the same single--particle state in either the $m=1$ or the $m=-1$ subspace \cite{Hu, DeMarco}. As the many--body ground state is two--fold degenerate, in the following we will consider a condensate formed in the $m=1$ subspace. For $\Omega > \Omega_c$ and weak $g$ there is a condensation into the $m = 0$ state.
 
\begin{figure}[!th]  
\includegraphics[width=8cm]{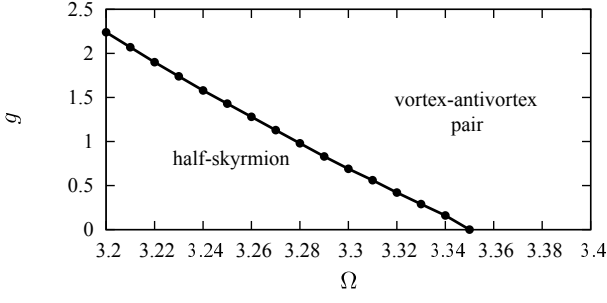}
\caption{Transition line between half--skyrmion and $m=0$ condensate, which was originally calculated in Ref.~\cite{Hu}.}
\label{Fig:Figgs}
\end{figure}

The total energy per particle of the condensed state with the order parameter $\left(\psi_1({\bf r}) \, \psi_2({\bf r}) \right)^T $ is given by
\begin{eqnarray}
 E_0 &=&  \int d{\bf r} \left[\left(\psi_1^* \psi_2^*\right) H_0 \left(\psi_1 \psi_2\right)^{T}\right.\nonumber\\
 &&\left. + \frac{1}{2}g |\psi_1|^4 +  \frac{1}{2}g |\psi_2|^4 + g |\psi_1|^2 |\psi_2|^2\right].
\end{eqnarray}
In order to find the ground state, we perform minimization of this functional with respect to $\psi_1({\bf r})$ and $\psi_2({\bf r})$. As usual, we introduce a chemical potential $\mu$ to enforce a normalization condition $\int d{\bf r}\left( |\psi_1({\bf r})|^2+|\psi_2({\bf r})|^2 \right) = 1.$ In the ground state, we have
\begin{eqnarray}
\mu \psi_1^0 &=& \left[\frac{p^2}{2}+\frac{r^2}{2}\left(1+\Omega^2\right)+g \left(|\psi_1^0|^2 + |\psi_2^0|^2 \right)\right]\psi_1^0\nonumber\\
&+& \frac{r^2 }{2}\Omega^2e^{-2 i \phi} \psi_2^0,
\label{eq:statgp1}\\
\mu \psi_2^0 &=& \left[\frac{p^2}{2}+\frac{r^2}{2}\left(1+\Omega^2\right)+g \left(|\psi_1^0|^2 + |\psi_2^0|^2 \right)\right]\psi_2^0 \nonumber\\
&+& \frac{r^2}{2}\Omega^2 e^{2 i \phi} \psi_1^0, 
\label{eq:statgp2}
\end{eqnarray}
where the chemical potential $\mu$ is given by
$ \mu =  \int d{\bf r} \left[\left(\psi_1^{0*} \,\psi_2^{0*}\right) H_0 \left(\psi_1^0 \,\psi_2^0\right)^{T}+ g \left(|\psi_1^0|^2 + |\psi_2^0|^2\right)^2\right].$ By comparing the ground--state energies of the condensed state in the two subspaces $m=0$ and $m=1$, it has been established that even at $\Omega <\Omega_c$ there is a transition into an $m=0$ condensate with increasing $g$, as shown in Fig.~\ref{Fig:Figgs}, which was originally calculated in Ref.~\cite{Hu}.

In order to learn about low--energy excitations of the condensed phase, we use the Bogoliubov approach. It can be performed on the operator level, or starting from the time--dependent Gross-Pitaevskii equation for $\psi_1({\bf r}, t) $ and $ \psi_2({\bf r}, t)$ \cite{Giorgini}:
\begin{eqnarray}
  i\frac{\partial \psi_1}{\partial t} &=& \left[\frac{p^2}{2}+\frac{r^2}{2}\left(1+\Omega^2\right)\right]\psi_1+ \frac{1}{2}\Omega^2 r^2 e^{-2 i \phi} \psi_2\nonumber\\ 
  &+& g |\psi_1|^2 \psi_1+g |\psi_2|^2 \psi_1, \label{eq:tdgp1}\\
  i \frac{\partial \psi_2}{\partial t} &=& \left[\frac{p^2}{2}+\frac{r^2}{2}\left(1+\Omega^2\right)\right]\psi_2 + \frac{1}{2}\Omega^2 r^2 e^{2 i \phi} \psi_1\nonumber\\
  &+& g |\psi_2|^2 \psi_2+g |\psi_1|^2 \psi_2.
  \label{eq:tdgp2}
\end{eqnarray}
In the following, we use the second approach.
%consider ground states that obey the symmetries of the initial Hamiltonian $\mathcal{H}_0$ discussed in the previous section.

Our first assumption is that the fluctuations $\delta \psi_1({\bf r}, t) $ and $ \delta \psi_2({\bf r}, t)$ around the ground state,
\begin{eqnarray}
 \psi_1({\bf r}, t)&\approx&\left[\psi_1^0({\bf r})+\delta \psi_1({\bf r}, t)\right] \exp(-i \mu t),\\
  \label{eq:fluctuations1}
 \psi_2({\bf r}, t)&\approx&\left[\psi_2^0({\bf r})+\delta \psi_2({\bf r}, t)\right] \exp(-i \mu t),
 \label{eq:fluctuations2}
\end{eqnarray}
 are weak. At the zeroth order in the fluctuations, from Eqs.~(\ref{eq:tdgp1})--(\ref{eq:tdgp2}) we recover Eqs.~(\ref{eq:statgp1})--(\ref{eq:statgp2}). By keeping terms of the first order, we derive a
 set of linear equations that describe the low--lying excitations of our system.
To decouple the equations further, we proceed in a standard way and introduce
\begin{eqnarray}
 \delta \psi_1({\bf r}, t) &=& u_1({\bf r}) \exp(-i \omega t)  + v_1^*({\bf r}) \exp(i \omega t),\\
  \delta \psi_2({\bf r}, t) &=& u_2({\bf r}) \exp(-i \omega t)  + v_2^*({\bf r}) \exp(i \omega t),
\end{eqnarray}
to obtain the generalized eigenproblem
\begin{eqnarray}
&& \omega\, u_1 = \left(\frac{p^2}{2}+\frac{r^2}{2}\left(1+\Omega^2\right)+2 g |\psi_1^0|^2+g |\psi_2^0|^2 -\mu\right) u_1\nonumber\\
 &+&\frac{r^2}{2}\Omega^2 e^{-2 i \phi} u_2 +  g (\psi_1^0)^2 v_1 + g \psi_1^0 \psi_2^{0*} u_2+g \psi_1^0 \psi_2^0 v_2,
 \label{eq:set1}\\
 &&- \omega\, v_1 = \left(\frac{p^2}{2}+\frac{r^2}{2}\left(1+\Omega^2\right)+2 g |\psi_1^0|^2+g |\psi_2^0|^2 -\mu\right) v_1\nonumber\\
 &+&\frac{r^2}{2}\Omega^2 e^{2 i \phi} v_2
 + g \left(\psi_1^{0*}\right)^2 u_1 + g \psi_1^{0*} \psi_2^0 v_2+g \psi_1^{0*} \psi_2^{0*} u_2,
  \label{eq:set2}\\
&& \omega\, u_2 = \left(\frac{p^2}{2}+\frac{r^2}{2}\left(1+\Omega^2\right)+g |\psi_1^0|^2+2 g |\psi_2^0|^2-\mu\right) u_2\nonumber\\
 &+&\frac{r^2}{2}\Omega^2  e^{2 i \phi} u_1+g (\psi_2^0)^2 v_2 + g \psi_1^0 \psi_2^0 v_1+g \psi_1^{0*} \psi_2^0 u_1,
  \label{eq:set3}\\
&& -\omega\, v_2 = \left(\frac{p^2}{2}+\frac{r^2}{2}\left(1+\Omega^2\right)+g |\psi_1^0|^2 +2 g |\psi_2^0|^2-\mu\right) v_2\nonumber\\
 &+&\frac{ r^2}{2}\Omega^2 e^{-2 i \phi}v_1 + g \psi_1^{0*} \psi_2^{0*} u_1+ g \psi_2^{0*} \psi_1^0 v_1+ g (\psi_2^{0*})^{2} u_2.
 \label{eq:set4}
\end{eqnarray}
In general, the resulting eigenvalues form pairs $-\omega_n, \omega_n$ and only positive frequencies correspond to physical excitations of the system.

To complement the Bogoliubov method, we numerically solve Eqs.~(\ref{eq:tdgp1})--(\ref{eq:tdgp2}) for different types of perturbations (\ref{eq:pertbm}) and (\ref{eq:pertdm}). For this purpose, the existing numerical codes for the two--dimensional time--dependent Gross-Pitaevskii equations \cite{Muruganandam, Vudragovic, Kumar, Loncar, Sataric, Young} have been modified to include the  spin--angular momentum coupling from Eq.~(\ref{eq:h0}).

 \section{Results}
 
 In this section we present and discuss excitation spectra and dynamical responses to perturbations (\ref{eq:pertbm}) and (\ref{eq:pertdm}) of the half--skyrmion and the $m=0$ condensate.
 
 \subsection{Half--skyrmion state}
 
 We first consider the case of $\Omega <\Omega_c$ and weak interaction $g$, where all bosons condense into $m=1$  state. 
 By inspecting Eqs.~(\ref{eq:set1}--\ref{eq:set4}) for the $\phi$--dependent terms, where we take into account a non--trivial $\phi$ dependence of the order parameters $\psi_1({\bf r}) $ and $\psi_2({\bf r}) $, we can infer that the solution can be cast in the form
  \begin{figure}[!th]
   \includegraphics[width=8cm]{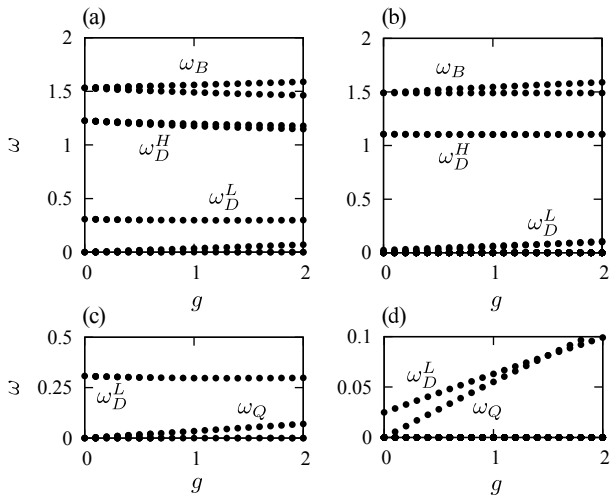}
\caption{Excitation spectra of half--skyrmion state for:  (a), (c) $\Omega = 2$ and (b), (d) $\Omega = 3.2$. Results obtained by the Bogoliubov approach.}
\label{Fig:Figsp}
\end{figure}
\begin{equation}
 \left(\begin{array}{c}
  u_1({\bf r})\\
  v_1({\bf r})\\
  u_2({\bf r})\\
  v_2({\bf r})
 \end{array}\right)=\sum_m 
  \left(\begin{array}{c}
u_1^{m-1}(r)r^{|m-1|}\exp(i (m-1) \phi)\\
v_1^{m-1}(r)r^{|m-1|}\exp(i (m-1) \phi)\\
u_2^{m+1}(r)r^{|m+1|}\exp(i (m+1) \phi)\\
v_2^{m-3}(r)r^{|m-3|}\exp(i (m-3) \phi)\\
 \end{array}\right).
\end{equation}
The explicit form of the matrices, that are diagonalized, are given in Appendix \ref{Sec:Appendix2}. The obtained spectrum shares many features with the noninteracting spectrum presented in Fig.~\ref{Fig:Fig0}(b), but it also exhibits important differences.

\begin{figure}[!bh]
  \includegraphics[width=\linewidth]{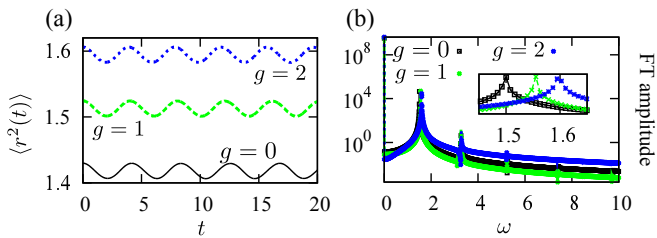}
\caption{Breathing mode oscillations in half--skyrmion phase: (a) $\langle r^2(t) \rangle $ versus $t$  and (b) corresponding Fourier transform.  From the inset we observe increase of the breathing mode frequency  with $g$.  Motion is induced by changing harmonic trap potential as $\frac{r^2}{2} \rightarrow 1.01\frac{r^2}{2} $, $\Omega = 3.2$.}
\label{Fig:FigbmM1}
\end{figure}

Excitation frequencies as a function of the interaction strength $g$ are plotted in Fig.~\ref{Fig:Figsp}(a) for $\Omega = 2$ and in Fig.~\ref{Fig:Figsp}(b) for $\Omega = 3.2$. The lowest excitation that does not change the relevant quantum number of the ground state is the breathing mode and its frequency increases for several percent with $g$. This is also confirmed by solving  Eqs.~(\ref{eq:tdgp1})--(\ref{eq:tdgp2}) in order to obtain  $\langle r^2(t)\rangle $,  as shown in Fig.~\ref{Fig:FigbmM1}(a), and then inspecting corresponding Fourier transforms, Fig.~\ref{Fig:FigbmM1}(b).

  \begin{figure}[!th]
   \includegraphics[width=\linewidth]{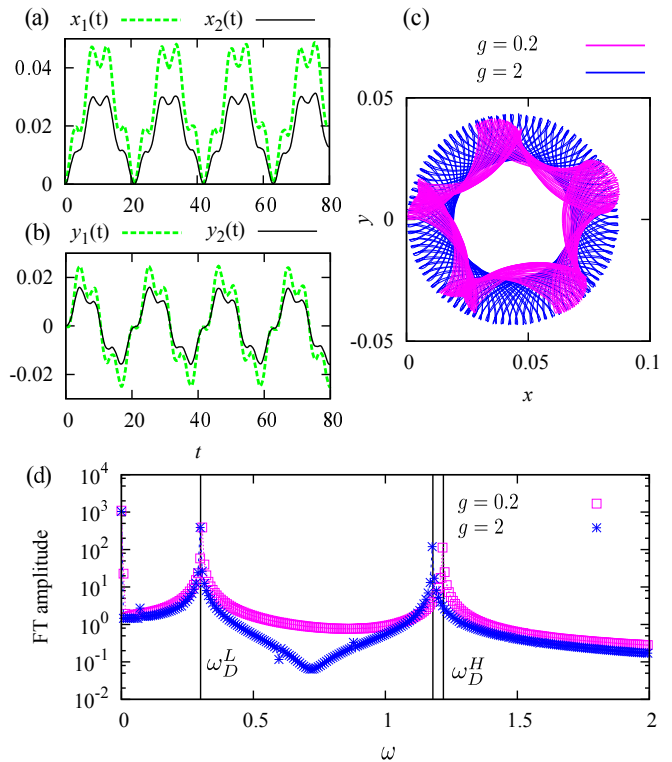}
\caption{ Dipole mode oscillations of half--skyrmion state in interacting case for $\Omega = 2$. Motion is induced by shifting harmonic trap bottom for $\delta x = 0.02$. In (a) and (b) $g = 1$. In (c)   motion of the center of mass, $y(t)$ versus $x(t)$, is plotted. In (d) vertical lines give results for $\omega_D^L$ and $\omega_D^H$ obtained using the  Bogoliubov method, and dots represent Fourier transform of $x(t)$.}
\label{Fig:FigdmM1Omega2}
\end{figure}

The most obvious difference with respect to the non--interacting spectrum is that the quadrupole mode is now gapped: at finite interaction $g$ it costs some energy to move a  particle from the half--skyrmion $m=1$ condensate into the $m=-1$ state, 
see Fig.~\ref{Fig:Figsp}(c) and  Fig.~\ref{Fig:Figsp}(d). This is directly reflected onto the dipole mode oscillations, that take place in the $xy$--plane for the half--skyrmion state. For $\Omega = 2$ both $\omega_D^L$ and $\omega_D^H$ are only weakly affected by $g$, however the fact that the quadrupole mode is gapped means that now a simpler perturbation theory applies. 
In the first order of this theory in $\delta x $ the center--of--mass motion is given by
 \begin{eqnarray}
  \langle x (t) \rangle &\approx& \frac{\delta x}{2}\left(\frac{I_{10}^2}{\omega_D^L}\cos \omega_D^L t+\frac{I_{12}^2}{\omega_D^H}\cos \omega_D^H t\right)+\mathrm{const},\\
    \langle y (t) \rangle &\approx& \frac{\delta x}{2}\left(\frac{I_{10}^2}{\omega_D^L}\sin \omega_D^L t+\frac{I_{12}^2}{\omega_D^H}\sin \omega_D^H t\right),
 \end{eqnarray}
 where the values of $I_{10}$ and $I_{12}$ can be roughly approximated by using the non--interacting eigenstates from Eq.~(\ref{eq:phim}) as $I_{10} = \int_0^{\infty} dr r^2 f_0^*(r) \left[f_1(r) - g_1(r)\right]$ and $I_{12} = \int_0^{\infty} dr r^2  \left[f_1^*(r) f_2(r)+ g_1^*(r)g_2(r)\right]$. 
 \begin{figure}[!th]
    \includegraphics[width=\linewidth]{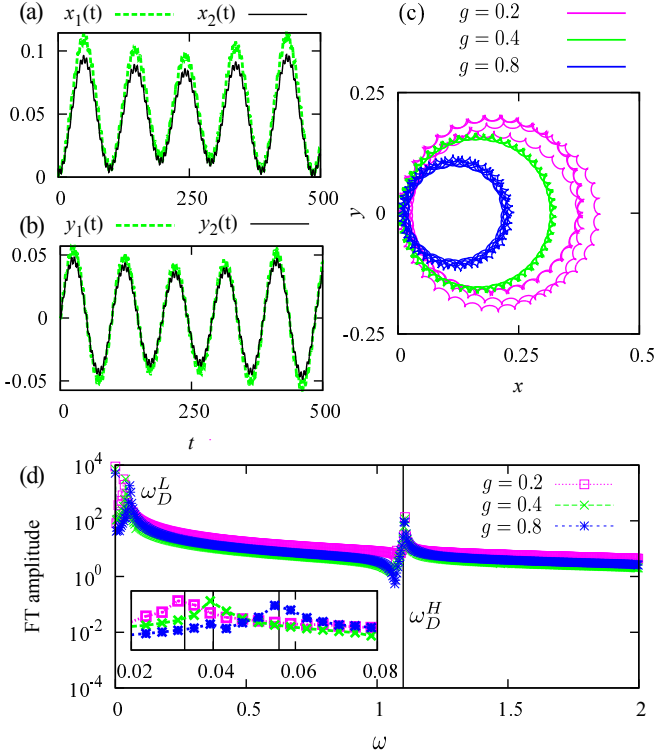}
\caption{Dipole mode oscillations of half--skyrmion state  in interacting case for $\Omega = 3.2$. Motion is induced by shifting the harmonic trap bottom for $\delta x = 0.01$. In (a) and (b) $g = 1$.  In (c)   motion of the center of mass, $y(t)$ versus $x(t)$, is plotted. The trajectory radius gets smaller with increasing $g$. In (d) vertical lines give results for $\omega_D^L$ (in the inset) and $\omega_D^H$ (in the main panel) obtained using the  Bogoliubov method, and dots represent Fourier transform of $x(t)$.  }
\label{Fig:FigdmM1}
\end{figure}
 In Fig.~\ref{Fig:FigdmM1Omega2}(c) we see how the pattern in the $xy$--plane becomes regular and symmetric as $g$ is changed from $g = 0.2$ to $g=2$. The two bosonic components oscillate in--phase in both directions, see Figs.~\ref{Fig:FigdmM1Omega2}(a) and \ref{Fig:FigdmM1Omega2}(b). Results of the Bogoliubov approach, which are captured by Eqs.~(\ref{eq:set1})--(\ref{eq:set4}), match quite well to the numerical data obtained from direct numerical simulations of Eqs.~(\ref{eq:tdgp1})--(\ref{eq:tdgp2}), see Fig.~\ref{Fig:FigdmM1Omega2}(d).

Effects of interactions are more prominent close to $\Omega_c$. In this case the frequency $\omega_D^L$ exhibits a strong increase with $g$, as is depicted in Fig.~\ref{Fig:Figsp}(d).
In Figs.~\ref{Fig:FigdmM1}(a) and \ref{Fig:FigdmM1}(b) we see that the oscillations are still as strong as for $g = 0$, but the pattern is regular, compare Fig.~\ref{Fig:FigdmM1}(c) with Fig.~\ref{Fig:FigdmM1g0}(d). As the frequency $\omega_D^L$ gets larger, the induced oscillation amplitude gets weaker and the induced frequency is less affected by the shift of the trap $\delta x $. In this case, the frequency $\omega_D^H $ is found to be almost independent of $g$, see Figs.~\ref{Fig:Figsp}(b) and \ref{Fig:FigdmM1}(d).

\subsection{Vortex--antivortex pair}

\begin{figure}[!th]
 \includegraphics[width=\linewidth]{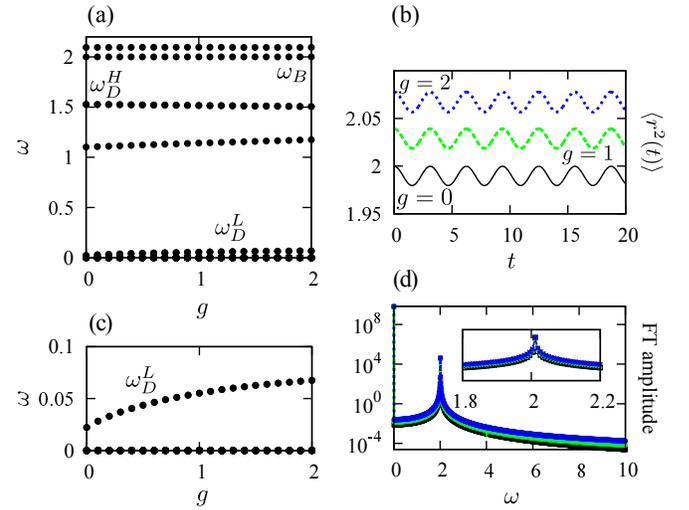}
\caption{Excitation spectra of the $m = 0$ state for  $\Omega = 3.5$, (a) and (c). Breathing mode oscillations for the $m = 0$ solution: (b) $\langle r^2(t) \rangle $ versus $t$  and (d) corresponding Fourier transform.  Motion is induced by changing harmonic trap potential as $\frac{r^2}{2} \rightarrow 1.01\frac{r^2}{2} $ for $\Omega = 3.5$.}
\label{Fig:FigspM0}
\end{figure}
In a similar way we proceed in the case of $\Omega > \Omega_c$, where the bosons condense in  the $m = 0$ state. The solution of Eqs.~(\ref{eq:set1})--(\ref{eq:set4}) can now be cast in the form:
\begin{equation}
 \left(\begin{array}{c}
  u_1({\bf r})\\
  v_1({\bf r})\\
  u_2({\bf r})\\
  v_2({\bf r})
 \end{array}\right)=\sum_m 
  \left(\begin{array}{c}
u_1^{m-1}(r)r^{|m-1|}\exp(i (m-1) \phi)\\
v_1^{m+1}(r)r^{|m+1|}\exp(i (m+1) \phi)\\
u_2^{m+1}(r)r^{|m+1|}\exp(i (m+1) \phi)\\
v_2^{m-1}(r)r^{|m-1|}\exp(i (m-1) \phi)\\
 \end{array}\right).
\end{equation}
Excitation frequencies as a function of the interaction strength $g$ are plotted in Fig.~\ref{Fig:FigspM0}. As anticipated in Sec.~II, the breathing mode frequency of the $m = 0$ state is independent of $g$  and at the mean--field level we have  $\omega_B = 2$ \cite{Pitaevskii}.

\begin{figure}[!th]
\includegraphics[width=\linewidth]{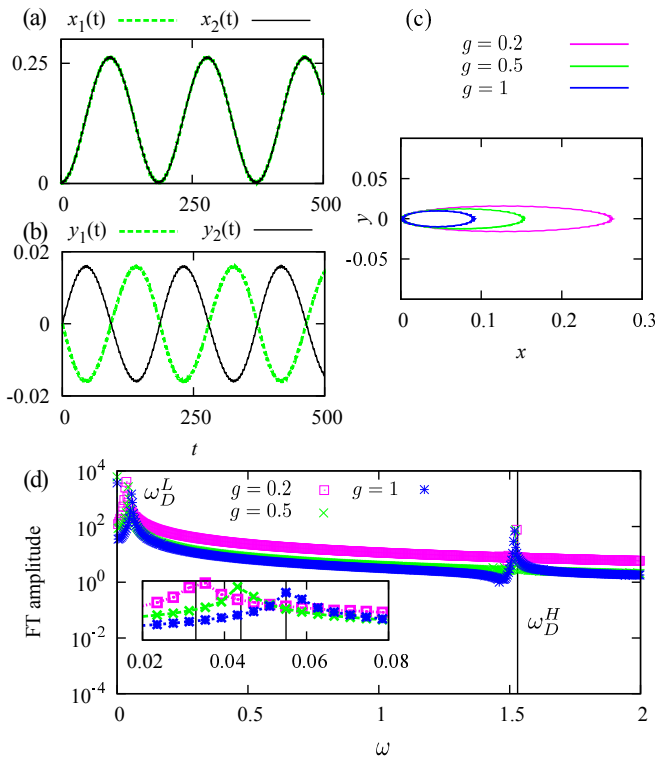}
\caption{Dipole mode oscillations of the $m=0$ solution in interacting case for $\Omega = 3.5$. Motion is induced by shifting harmonic trap bottom for $\delta x = 0.01$. In (a) and (b) $g = 0.2$. In (c) trajectory of the center of mass of a single bosonic component, $y_1(t)$ versus $x_1(t)$, is plotted. In (d) vertical lines give results for $\omega_D^L$ (in the inset) and $\omega_D^H$ (in the main panel) obtained using Bogoliubov method and dots represent Fourier transform of $x(t)$. }
\label{Fig:FigdmM0}
\end{figure}

In the dipole mode oscillations, the two bosonic components exhibit an out--of--phase oscillation in $y$ direction, see Fig.~\ref{Fig:FigdmM0}(b), and consequently the center of mass only oscillates in $x$ direction with the frequency $\omega_D^L$
that exhibits an increase with $g$, see Fig.~\ref{Fig:FigspM0}(b). The trajectory of the center of mass of each of the components is given by an ellipse, which is strongly elongated in $x$ direction, see Fig.~\ref{Fig:FigdmM0}. A much weaker effect of $g$ is observed in $\omega_D^H$, that is quite close to the numerical resolution of the applied methods.

\subsection{Discussion}

The ground state mean--field calculations indicate a first--order phase transition from a half--skyrmion state into $m=0$ condensate with increasing $g$ at $\Omega < \Omega_c$ and $g = g_c$ \cite{Hu} as shown in Fig.~\ref{Fig:Figgs}. Based on the Bogoliubov analysis we find that this $m = 0$ state is dynamically unstable for $\Omega < \Omega_c$ at $g > g_c$ as it exhibits an imaginary excitation frequency. The results of the numerical simulations of Eqs.~(\ref{eq:tdgp1})--(\ref{eq:tdgp2}) also show a nonlinear behavior in this regime, such as mode coupling and the generation of higher harmonics. 
%This aspect is somewhat unexpected, as well as the absence of a mode softening as we approach the transition line. 
One way to resolve this issue is to use a method that is an alternative to the mean--field calculation, such as exact diagonalization. Although this method suffers from conceptual limitations in higher dimensions, if the two--body interactions are described by a contact potential (Dirac delta function) \cite{Saarikoski, Esry}, we have implemented it with a finite energy cutoff, as described in Ref.~\cite{Haugset}. In particular, we perform a simplified diagonalization study for $\Omega$ close to $\Omega_c$ by taking into account only the three nearly degenerate noninteracting eigenstates. This analysis is sufficient to discuss the change in the ground state and the two lowest excitations $\omega_Q$ and $\omega_D^L$. 
%It provides us with an indication regarding the effects that we do not capture within the mean--field treatment of the model Hamiltonian (\ref{eq:inth}).

\begin{figure}[!th]
\includegraphics[width=8cm]{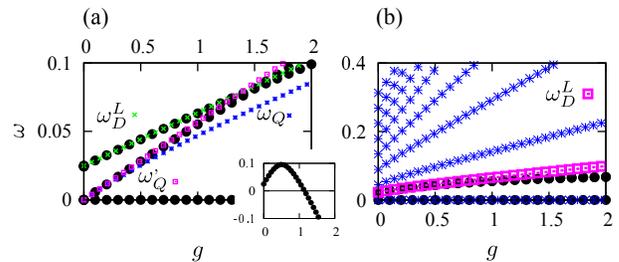}
\caption{Comparison of Bogoliubov analysis (black dots) and simplified diagonalization: (a) $\Omega = 3.2$ and (b) $\Omega = 3.5$. Inset gives energy difference of  $m = 0$ state, which turns out to exhibit a condensate fraction significantly smaller than 1, and $m = 15$ state, which corresponds to a half--skyrmion condensate.}
\label{Fig:Figed}
\end{figure}

A comparison of the results obtained by the simplified diagonalization and by the  Bogoliubov method is given in Fig.~\ref{Fig:Figed} for $N_p =15$ particles used in the diagonalization, where we see that the two methods show good agreement in $\omega_D^L$ in both phases. However, the frequency $\omega_Q$ is overestimated in the Bogoliubov analysis. This can be understood as follows: when performing a diagonalization, the lowest--lying state in the sector $m = N_p-2$ is a linear combination of states $|-11^{Np-1}\rangle $ and $|1^{Np-2}00\rangle $. 
However, the frequency $\omega_Q$, that we obtained using the Bogoliubov method,  corresponds much better to the energy expectation value of $|-11^{Np-1}\rangle $, from which we subtract $E_0$, as it neglects the two--particle excitations. An effect of similar origin is found for the $m=0$ condensate at $\Omega > \Omega_c$, where we find a series of two--particle excitations $|0^{N_p}\rangle \rightarrow |0^{N_p-2}1-1\rangle \rightarrow |0^{N_p-4}11-1-1\rangle $ with the same quantum number as the ground state that we do not capture using the  Bogoliubov method, see Fig.~\ref{Fig:Figed}(b). In the inset of Fig.~\ref{Fig:Figed}(a) we plot the energy difference between the two competing states for $\Omega = 3.2$. We find that  the transition from the half--skyrmion condensate to $m=0$ state occurs  at a lower value of $g$ compared to the mean--field prediction, and that the $m=0$ state obtained in this way has a condensate fraction substantially lower then $1$. For this reason in the region of the phase diagram $\Omega<\Omega_c, g>g_c $ beyond--mean--field effects become important. 
%Moreover, we infer that the change in the ground state is not directly reflected in   $\omega_Q$ and $\omega_D^L$, which agrees with our earlier findings. 

\section{Conclusions}
Motivated by ongoing experimental efforts to realize and probe new quantum states, we have investigated the breathing mode and the dipole mode oscillations of the half--skyrmion bosonic condensed state. These excitations are routinely used in the experiments and we find that both of them distinguish the half--skyrmion phase from a competing $m=0$ state. In particular, the breathing mode frequency of the half--skyrmion state depends on the spin--orbit coupling  and interaction strength, while it takes a universal value in the $m=0$ state at the classical level. As a response to the sudden shift of the harmonic trap, a center of mass of a half--skyrmion state exhibits a peculiar motion in the $xy$--plane that involves the two dominant excitation frequencies $\omega_D^L$ and $\omega_D^H$. In the non--interacting limit, the degeneracy of the $m = 1$ half--skyrmion with $m=-1$ state leads to complex motion patterns. Weak repulsive interactions make the quadrupole mode gapped and lead to simpler and symmetric patterns. These effects of interactions are stronger closer to the transition point between the two phases, where they prominently enhance the frequency $\omega_D^L$.

In future work, we plan to address bosonic excitations for spin--asymmetric interactions as well as to treat interactions for spin--orbit coupled system in more detail \cite{Gopalakrishnan}. Another interesting direction would be to investigate the role of disorder \cite{Krumnow, Nikolic, Ghabour, Mardonov, Khellil1, Khellil2}, or the phenomenon of Faraday waves \cite{Nicolin, Balaz, Balaz2}  in this type of systems.

\section{Acknowledgements}
The authors thank Axel Pelster for useful discussions.
This work was supported by the Ministry of Education, Science, and Technological Development of the Republic of Serbia under projects ON171017 and OI1611005, and by the European Commission under H2020 project VI-SEEM, Grant No. 675121. Numerical simulations were performed on the PARADOX supercomputing facility at the Scientific Computing Laboratory of the Institute of Physics Belgrade.

\appendix
\begin{widetext}
\section{Perturbation theory for nearly degenerate states}
\label{Sec:Appendix1}
To describe the lowest excitation frequencies, we consider the lowest lying states of the Hamiltonian (\ref{eq:h0}), given in Eq.~(\ref{eq:phim}), for $m = -1,0,1 $. The states $m=\pm 1$ are degenerate and the state $m=0$ is close in energy, see Fig.~\ref{Fig:Fig0}.
In the lowest order of the perturbation theory, the relevant part of the perturbed Hamiltonian (\ref{eq:pertdm}) can be approximated by
\begin{equation}
 H_{\mathrm{pert}}^{\mathrm{red}} = \left(\begin{array}{ccc}
                         a & c/2 & 0\\
                         c/2 & b & -c/2\\
                         0 & -c/2 & a
                        \end{array}\right),
\label{eq:hpert}                        
\end{equation}
where $a = E_{0}^{m=-1} = E_0^{m=1}$, $b  =E_0^{m=0} $, $c = \delta x I_{10} = \delta x \,\int_0^{\infty} dr \,r^2 f_0^*(r) \left[f_1(r) - g_1(r)\right]$, 
%where $I_{10} = \langle 0 | x \otimes \mathcal{I}_2|1\rangle $ 
and integrals over the angle $\phi$ have already been performed. Functions $f_0(r)$, $f_1(r)$ and $g_1(r)$ are defined in Eq.~(\ref{eq:phim}). For completeness, other relevant operators in this subspace are approximated by
\begin{eqnarray}
 J_z^{\mathrm{red}}=\left( \begin{array}{ccc}
             -1 &0 &0\\
             0 & 0 &0\\
             0&0&1
            \end{array}\right),\quad
x^{\mathrm{red}} \otimes \mathcal{I}_2 = \left( \begin{array}{ccc}
             0 & -d/2 & 0\\
             -d/2 & 0 & d/2\\
             0 & d/2 & 0
            \end{array}
\right),\quad
y^{\mathrm{red}} \otimes \mathcal{I}_2= \left(\begin{array}{ccc}
           0 & i d/2 & 0\\
           -i d/2 & 0 & -i d/2\\
           0 & i d/2 &  0
          \end{array}
\right),
\end{eqnarray}
where $d = I_{10}$.
The eigensystem of $H_{\mathrm{pert}}^{\mathrm{red}}$ is given by
\begin{equation}
 E_1 = a,\quad E_2 = \frac{a+b-z}{2}, \quad E_3 = \frac{a+b+z}{2},
\end{equation}
\begin{equation}
 v_1=\frac{1}{\sqrt{2}}\left(\begin{array}{c} 1\\0\\1\end{array}\right), \quad v_2 = \frac{1}{\sqrt{n_2}}\left(\begin{array}{c} -1\\ \frac{z-\omega_D^L}{c}\\1\end{array}\right), \quad v_3 = \frac{1}{\sqrt{n_3}}\left(\begin{array}{c} -1\\ -\frac{z+\omega_D^L}{c}\\1\end{array}\right),
\end{equation}
where $\omega_D^L = b-a$, $z=\sqrt{{\omega_D^L}^2+2 c^2}$, $n_2 = 2 z (z-\omega_D^L)/c^2 $, $n_3 = 2 z(z+\omega_D^L)/c^2 $.

First we consider the case when the system is initially prepared in the half -- skyrmion configuration $|\psi(t = 0) \rangle = \left(0\quad 0 \quad1\right)^T $. With this initial condition, we have
\begin{equation}
 |\psi(t)\rangle \approx \frac{1}{2} \left(\begin{array}{c} 1\\ 0\\ 1\end{array} \right) e^{-i E_1 t}+\frac{1}{n_2} \left(\begin{array}{c} -1\\ \frac{z-\omega_D^L}{c}\\ 1\end{array} \right) e^{-i E_2 t}+\frac{1}{n_3} \left(\begin{array}{c} -1\\ -\frac{z+\omega_D^L}{c}\\ 1\end{array} \right) e^{-i E_3 t}.
\end{equation}
From the last expression we can find all expectation values $\langle O(t) \rangle = \langle \psi(t)| O |\psi(t) \rangle $. We start from 
\begin{equation}
\langle J_z(t) \rangle \approx \frac{2}{n_2} \cos \left(E_2 - E_1 \right)t + \frac{2}{n_3} \cos \left(E_3 - E_1 \right)t.
\end{equation}
From Eq.~(\ref{eq:ymotion}) it follows directly
\begin{eqnarray}
\langle y(t) \rangle &\approx&\frac{2d}{c}\frac{E_2 - E_1}{n_2} \sin \left(E_2 - E_1 \right)t + \frac{2d}{c}\frac{E_3-E_1}{n_3} \sin \left(E_3 - E_1 \right)t \nonumber\\
&=& \frac{\delta x \, I_{10}^2}{2 \sqrt{{\omega_D^L}^2+2 \left(\delta x I_{10} \right)^2}}\left[\sin\left(\frac{\sqrt{{\omega_D^L}^2+2 \left(\delta x I_{10}\right)^2}-\omega_D^L}{2}\right) t+\sin\left(\frac{\sqrt{{\omega_D^L}^2+2 \left(\delta x I_{10}\right)^2}+\omega_D^L}{2}\right) t\right].
\label{eq:yoft}
\end{eqnarray}
When calculating the expectation value of $x ^{\mathrm{red}} $, we first note that $x ^{\mathrm{red}} v_1 = 0$, $v_1^T x ^{\mathrm{red}} v_{2,3}=0 $. From here it follows that the expectation value will oscillate with the frequency $E_3 - E_2$.
The straightforward calculation yields
\begin{equation}
 \langle x(t) \rangle \approx  \frac{\delta x \, I_{10}^2 \,\omega_D^L}{2 \left({\omega_D^L}^2+2 \left(\delta x I_{10} \right)^2\right)}\left[1-\cos\sqrt{{\omega_D^L}^2+2 \left(\delta x I_{10} \right)^2} t\right].
 \label{eq:xoft}
\end{equation}
Results captured by Eqs.~(\ref{eq:xoft}) and (\ref{eq:yoft}) are presented in Fig.~\ref{Fig:FigdmM1g0}, where we see that they reasonably agree with the full numerical calculation.

Next we consider the time evolution of the vortex--antivortex pair $|\psi(t = 0) \rangle = \left(0\quad 1 \quad0\right)^T $. In this case
\begin{equation}
 |\psi(t)\rangle \approx \frac{z-\omega_D^L}{c n_2} \left(\begin{array}{c} -1\\ \frac{z-\omega_D^L}{c}\\ 1\end{array} \right) e^{-i E_2 t}-\frac{z+\omega_D^L}{c n_3} \left(\begin{array}{c} -1\\ -\frac{z+\omega_D^L}{c}\\ 1\end{array} \right) e^{-i E_3 t}.
\end{equation}
As the perturbation couples the $m=0$ state symmetrically to $\pm m$ states, we find $\langle J_z(t) \rangle = 0$ and the motion occurs only in the $x$ direction, where we recover Eq.~(\ref{eq:xoft}).

\section{Explicit form of Bogoliubov equations}
\label{Sec:Appendix2}

For a half--skyrmion ground state we rewrite linearized Eqs.~(\ref{eq:set1})--(\ref{eq:set4}) in the form of the  eigenproblem of the matrix   $\mathcal{H}_{Bg, hs}$,
 \begin{equation}
   \mathcal{H}^m_{Bg, hs}  \left(\begin{array}{c}
  u_1^{m-1}({\bf r})\\
  v_1^{m-1}({\bf r})\\
  u_2^{m+1}({\bf r})\\
  v_2^{m-3}({\bf r})
 \end{array}\right)=\omega  \left(\begin{array}{c}
  u_1^{m-1}({\bf r})\\
  v_1^{m-1}({\bf r})\\
  u_2^{m+1}({\bf r})\\
  v_2^{m-3}({\bf r})
 \end{array}\right),
 \end{equation}
where
 \begin{equation}
  \mathcal{H}_{Bg, hs} =  \mathcal{H}^0_{Bg, hs} + \mathcal{H}^g_{Bg, hs}-\mathcal{D}_{\mu},
 \end{equation}
  \begin{equation}
  \mathcal{H}^{0, m}_{Bg, hs} = \left(\begin{array}{cccc}
                   \mathcal{H}_0^{m-1} +  g |\psi_1^0|^2 + g |\psi_2^0|^2 & 0 & \frac{1}{2}\Omega^2 r^2 q_{13}(r) & 0 \\
                   0 & -\mathcal{H}_0^{m-1} - g |\psi_1^0|^2 - g |\psi_2^0|^2  & 0 & -\frac{1}{2}\Omega^2 r^2 q_{24}(r)\\
                  \frac{1}{2}\Omega^2 r^2q_{31}(r) & 0 & \mathcal{H}_0^{m+1} +  g |\psi_2^0|^2 + g |\psi_1^0|^2&0\\
                  0 &\ -\frac{1}{2}\Omega^2 r^2 q_{41}(r) &  0& -\mathcal{H}_0^{m-3} - g |\psi_2^0|^2 - g |\psi_1^0|^2 
                  \end{array}
                  \right),
 \end{equation}
 and
  \begin{equation}
  \mathcal{H}^{g, m}_{Bg, hs} = g\,\left(\begin{array}{cccc}
                     |\psi_1^0|^2  &  (\psi_1^0)^2 &   \psi_1^0 \chi_2^0 q_{13}(r) &  \psi_1^0 \chi_2^0 q_{14}(r) \\
                   - (\psi_1^0)^2 & -  |\psi_1^0|^2   & -  \psi_1^0 \chi_2^0 q_{23}(r) & -  \psi_1^0 \chi_2^0 q_{24}(r)\\
                   \psi_1^0 \chi_2^0 q_{31}(r)&  \psi_1^0 \chi_2^0 q_{31}(r)&   |\psi_2^0|^2 & (\chi_2^0)^2 q_{34}(r)\\
                   - \psi_1^0 \chi_2^0 q_{41}(r) & - \psi_1^0 \chi_2^0 q_{41}(r) &  - (\chi_2^0)^2 q_{43}(r)& -  |\psi_2^0|^2 
                  \end{array}
                  \right),\quad   \mathcal{D}_{\mu} =\mu \left(\begin{array}{cccc}
                             1&0&0&0\\
                             0&-1&0&0\\
                             0&0&1&0\\
                             0&0&0&-1
                            \end{array}\right).
 \end{equation}
We have introduced the following functions
\begin{eqnarray}
&& \chi_2(r) = \psi_2({\bf r}) \exp\left(-2 i \phi\right),\\
 &&q_{13}(r)=r^{|m+1|-|m-1|},\quad q_{14}(r) = r^{|m-3|-|m-1|},\quad  q_{23}(r)=r^{|m+1|-|m-1|},\quad q_{24}(r) = r^{|m-3|-|m-1|},\nonumber\\
 &&q_{31}(r) = r^{|m-1|-|m+1|},\quad   q_{34}(r) = r^{|m-3|-|m+1|},\quad q_{41}(r) = r^{|m-1|-|m-3|},\quad q_{43}(r) = r^{|m+1|-|m-3|},
 \end{eqnarray}
and $\mathcal{H}_0^{m} = -\frac{1}{2}\left(\frac{2 |m|+1}{r}\frac{d}{dr}+\frac{d^2}{dr^2}\right)+\frac{1}{2} \left(1+\Omega^2\right) r^2 $. 

In a similar way we proceed in the case of $m=0$ ground state:
% \begin{equation}
%  \left(\begin{array}{c}
%   u_1({\bf r})\\
%   v_1({\bf r})\\
%   u_2({\bf r})\\
%   v_2({\bf r})
%  \end{array}\right)=\sum_m 
%   \left(\begin{array}{c}
% u_1^{m-1}(r)r^{|m-1|}\exp(i (m-1) \phi)\\
% v_1^{m+1}(r)r^{|m+1|}\exp(i (m+1) \phi)\\
% u_2^{m+1}(r)r^{|m+1|}\exp(i (m+1) \phi)\\
% v_2^{m-1}(r)r^{|m-1|}\exp(i (m-1) \phi)\\
%  \end{array}\right)
% \end{equation}
 \begin{equation}
  \mathcal{H}_{Bg, m0} =  \mathcal{H}^0_{Bg, m0} + \mathcal{H}^g_{Bg, m0}-\mathcal{D}_{\mu},
 \end{equation}
 with
  \begin{equation}
  \mathcal{H}^{0, m}_{Bg, m0} = \left(\begin{array}{cccc}
                   \mathcal{H}_0^{m-1} +  g |\psi_1^0|^2 + g |\psi_2^0|^2 & 0 & \frac{1}{2}\Omega^2 r^2 h(r) & 0 \\
                   0 & -\mathcal{H}_0^{m+1} - g |\psi_1^0|^2 - g |\psi_2^0|^2  & 0 & -\frac{1}{2}\Omega^2 r^2 e(r)\\
                  \frac{1}{2}\Omega^2 r^2 e(r) & 0 & \mathcal{H}_0^{m+1} +  g |\psi_2^0|^2 + g |\psi_1^0|^2&0\\
                  0 &\ -\frac{1}{2}\Omega^2 r^2 h(r) &  0& -\mathcal{H}_0^{m-1} - g |\psi_2^0|^2 - g |\psi_1^0|^2
                  \end{array}
                  \right),
 \end{equation}
 and
  \begin{equation}
  \mathcal{H}^{g, m}_{Bg, m0} =  g\,|\psi_1^0|^2\left(\begin{array}{cccc}
                    1  &   h(r)&  - h(r) & -1   \\
                   - e(r)& - 1  &  1  &  e(r)\\
                  -e(r)& -1  &  1  & e(r)\\
                   1 & h(r)&  -h(r)& - 1
                  \end{array}
                  \right),
 \end{equation}
where $h(r)=r^{|m+1|-|m-1|}$, $e(r)=r^{|m-1|-|m+1|}$.

\end{widetext}

 \end{document}